# Observation of electrical high-harmonic generation

Xiaozhou Zan[1,2,10], Ming Gong[3,10], Zitian Pan[1,2,10], Haiwen Liu[4,5], Jingwei Dong[1,2], Jundong Zhu[1,2], Le Liu[1,2], Yanbang Chu[1,2], Kenji Watanabe[6], Takashi Taniguchi[7], Dongxia Shi[1,2], Wei Yang[1,2,8*], Luojun Du[1,2*], Xin-Cheng Xie[3,5,9*], and Guangyu Zhang[1,2,8*]

[1]*Beijing National Laboratory for Condensed Matter Physics and Institute of Physics, Chinese Academy of Sciences, Beijing100190, China.*
[2]*School of Physical Sciences, University of Chinese Academy of Sciences, Beijing100190, China.*
[3]*International Center for Quantum Materials, School of Physics, Peking University, Beijing 100871, China*
[4]*Center for Advanced Quantum Studies, Department of Physics, Beijing Normal University, Beijing, China*
[5]*Institute for Nanoelectronic Devices and Quantum Computing, Fudan University, Shanghai 200433, China*
[6]*Research Center for Electronic and Optical Materials, National Institute for Materials Science, 1-1 Namiki, Tsukuba 305-0044, Japan*
[7]*Research Center for Materials Nanoarchitectonics, National Institute for Materials Science, 1-1 Namiki, Tsukuba 305-0044, Japan*
[8]*Songshan Lake Materials Laboratory, Dongguan, Guangdong523808, China.*
[9]*Hefei National Laboratory, Hefei 230088, China*
[10]These authors contributed equally: Xiaozhou Zan, Ming Gong, Zitian Pan.
*Corresponding authors: luojun.du@iphy.ac.cn; wei.yang@iphy.ac.cn; xcxie@pku.edu.cn; gyzhang@iphy.ac.cn

**High-harmonic generation (HHG), an extreme nonlinear effect, introduces an unprecedented paradigm to detect emergent quantum phases and electron dynamics inconceivable in the framework of linear and low-order nonlinear processes. As an important manifestation, the optical HHG (*o*-HHG) enables extraordinary opportunities to underpin attosecond physics. In addition to nonlinear optics, emerging nonlinear electric transport has been demonstrated recently and opens new paradigms to probe quantum phase transition, symmetry breaking, band geometrical and topological properties. Thus far, only electrical second-/third-harmonic generation in perturbative regime has been elucidated, while the electrical HHG (*e*-HHG) that can advance to extreme non-perturbative physics remains elusive. Here we report the observation of *e*-HHG up to 300th-order. Remarkably, the *e*-HHG shows a clear non-perturbative character and exhibits periodic oscillations with the reciprocal of driving current. Further, theoretical simulations corroborate the experiments, suggesting the contribution of singular distribution of Berry curvature near band edges. Our results demonstrate *e*-HHG in extreme nonlinear regime and may shed light on a plethora of exotic physics and applications, such as extreme non-equilibrium quantum phenomena, ultra-fast and coherent electrical signal generations and detections.**

High-order nonlinear effects—a research frontier of modern physics and materials science that can advance to the extreme non-perturbative/nonlinear regime—offer a new paradigm to explore the fundamental structural and dynamic properties of emergent quantum phases that are unattainable with linear and low-order nonlinear processes[1-4]. One spectacular example is the optical high-harmonic generation (*o*-HHG) in the exotic non-perturbative regime, which has enabled the birth of attosecond science and initiated a plethora of emergent technological applications[2,3,5,6], e.g., generation of extreme ultraviolet/soft x-ray coherent radiations[5,7], electronic structure probing[8,9], on-chip ultrafast nanophotonics[10], and multi-petahertz/lightwave electronics[11,12].

In addition to the well-studied nonlinear optical physics, emergent nonlinear electric transports have been demonstrated recently, characterized by a harmonic voltage signal nonlinearly dependent on the driving electric field[13-17]. Nonlinear transport measurements are typically driven at extremely low frequencies (e.g., ~100 Hz) and thus confine the underlying physics near the Fermi surface[16], in contrast to high-frequency nonlinear optics (~$10^{14}$ Hz) with contributions from both intraband and interband transitions[18]. The nonlinear harmonic transports open up an unprecedented paradigm to probe the quantum phase transition, symmetry breaking, spin-orbit interaction, magnetism, superconductivity, ferroelectricity, topological and geometrical properties[13-17,19-33]. For instance, second-order nonlinear Hall effects can go beyond the constraint of broken time-reversal symmetry of linear Hall responses and present new avenues to probe Berry curvature dipole and topological quantum phase transitions in inversion-asymmetric materials[13-16,33-36]. Meanwhile, third-order harmonic transports offer opportunities to underpin exotic band geometric property[19,37] and current-induced magnetization dynamics[38].

Although many promising progresses have been witnessed, the exploration of nonlinear electric transports has been limited to only second-/third-order harmonic generation with underlying physics confined in the perturbative regime. In this work, we demonstrate the electrical high-harmonic generation (*e*-HHG) with both even and odd harmonic orders up to 300th-order. In marked contrast to the perturbative nature of second-/third-order harmonic transport, *e*-HHG exhibits a clear non-perturbative character. More interestingly, an unexpected oscillation with the reciprocal of driving current is observed for *e*-HHG larger than 10th-order, and the oscillation frequency increases linearly with the *e*-HHG order. Theoretical simulations corroborate the experiments and unravel the potential mechanism of the observed e-HHG responses. The observation of *e*-HHG in non-perturbative and extreme nonlinear regime could afford unprecedented capabilities for exploring intriguing band topologic/geometric properties, Berry curvature/quantum metric multipoles, novel nanoelectronics and energy devices inaccessible by the lower order processes.

**Electrically tunable nonlinearity**

Figure 1a depicts the schematic diagram of the nonlinear harmonic transport measurements along the transverse direction on a standard Hall bar device. Briefly, an alternating current $I^\omega = I_0 \sin(\omega t)$ with an amplitude $I_0$ and an oscillating frequency $\omega$ is applied longitudinally along the device, and a transverse voltage $V_{xy}^{m\omega}$ with the frequency $m\omega$ is recorded by a lock-in amplifier, where the integer $m$ ($\geq 2$) is the harmonic order of nonlinear electric transports (see Methods for more details)[14,15,27]. Unless otherwise specified, all nonlinear harmonic transport measurements in the main text were performed along the transverse direction with a driving current of $\omega = 30.9$ Hz, at a base temperature of 1.7 K and zero magnetic field.

Figure 1b schematically shows the typical structure of hexagonal boron nitride (*h*-BN) encapsulated, dual-gated bilayer graphene devices, with graphite (Au) as the bottom (top) gate. Unless otherwise noted, all measurements were recorded from device #1, which shows high quality as evidenced by quantum oscillations and quantum Hall effect at rather low onset magnetic fields (please see Supplementary Note 1 for details). The dual-gate architecture allows us to tune the out-of-plane displacement field $D$ and carrier density $n$ independently. Here $D = (C_b V_{bg} - C_t V_{tg})/2\varepsilon_0$ and $n = (C_b V_{bg} + C_t V_{tg})/e$, where $e$ is the elementary charge, $\varepsilon_0$ denotes the vacuum permittivity, $C_b$ ($V_{bg}$) and $C_t$ ($V_{tg}$) are the geometrical capacitances per area (applied voltages) for bottom and top gates, respectively. Bilayer graphene is chosen here for two main reasons. First, bilayer graphene is a highly tunable platform, which inversion symmetry and valley-contrasting electron chirality can be continuously engineered by $D$[35,39-42]. Second, theoretical calculation predicts that bilayer graphene can show giant quantum nonlinearity, facilitates the possibility of *e*-HHG[40].

Figure 1c plots a colour map of the second-order harmonic voltage response $V_{xy}^{2\omega}$ against $D$ and $n$. $V_{xy}^{2\omega}$ scales linearly with the square of the longitudinal voltage response $V_{xx}^{\omega}$, and reverses its sign when the driving current direction and the second-order harmonic voltage probe connection are changed simultaneously (Fig. 1d), confirming the second-order nonlinear nature in perturbative regime and being highly consistent with previous reports[4,14,15]. Additionally, $V_{xy}^{2\omega}$ is independent of the driving frequency (please see Supplementary Note 2 for details), precluding the potential measurement artifacts, e.g., a spurious capacitive coupling effect[15,19]. Intriguingly, $V_{xy}^{2\omega}$ is highly tunable by $D$. $V_{xy}^{2\omega}$ almost vanish under zero $D$, while shows significant signals with increasing second-order nonlinear conductivity $\sigma_{xy}^{2\omega} = \sigma_0 \frac{V_{xy}^{2\omega} L^2}{(V_{xx}^{\omega})^2 W}$ as $D$ increases progressively (Fig. 1e), where $\sigma_0$ is the linear conductivity, $W$ is the device channel width and $L$ is the distance between the voltage probes along current path (see Methods for more details). Remarkably, the second-order harmonic response in bilayer graphene is quite strong. The maximum $\sigma_{xy}^{2\omega}$ can reach up to ~ 1.1 μm V$^{-1}$ Ω$^{-1}$ (Fig. 1e), three (two) orders larger than the values reported in WTe$_2$ (twisted WSe$_2$)[14,15,33], and competitive to the state-of-the-art results reported recently in graphene/*h*-BN and twisted bilayer graphene moiré superlattices[20,32]. This highlights the giant quantum nonlinearity of chiral Bloch electrons in gapped bilayer graphene, highly consistent with the recent theoretical calculations[40].

To further confirm the giant quantum nonlinearity in gapped bilayer graphene, we performed the third-order harmonic transport measurements. Figure 1f presents a colour plot of third-order harmonic voltage response $V_{xy}^{3\omega}$ as a function of $D$ and $n$. $V_{xy}^{3\omega}$ scales linearly with the cube of $V_{xx}^{\omega}$ (Fig. 1g), and keeps its sign unchanged when the driving current direction and signal probe connection are reversed simultaneously (Supplementary Note 3), implying the intrinsic third-order nonlinear nature in perturbative regime[4,19]. Figure 1h shows the $D$-dependent third-order nonlinear conductivity $\sigma_{xy}^{3\omega} = \sigma_0 \frac{V_{xy}^{3\omega} L^3}{(V_{xx}^{\omega})^3 W}$, increasing progressively with $D$. Astonishingly, the maximum $\sigma_{xy}^{3\omega}$ of bilayer graphene can reach an extremely large value of ~ 5000 μm$^2$ V$^{-2}$ Ω$^{-1}$, four and seven orders of magnitude larger than the values reported in $T_d$-MoTe$_2$ (~ 0.32 μm$^2$ V$^{-2}$ Ω$^{-1}$) and $T_d$-WTe$_2$

($\sim 4.9 \times 10^{-4}$ μm² V⁻² Ω⁻¹), respectively[19].

**Observation of *e*-HHG**

The extremely high quantum nonlinearity in bilayer graphene, as shown above, motivates us to explore the *e*-HHG. Figure 2a shows a representative frequency spectrum of bilayer graphene under $D = 0.3$ V/nm and $n = 3.5 \times 10^{10}$ $cm^{-2}$, excited by an alternating current with an amplitude of 500 nA at the oscillating frequency of 30.9 Hz. The frequency spectra under different driving frequencies are also included in the inset of Fig. 2a. Intriguingly, regularly spaced spectral peaks at the integer multiples of the driving frequency, corresponding to the 2nd- to 50th-order harmonic transport responses, are clearly visible, bearing striking resemblance to the *o*-HHG[2,3]. The existence of harmonic responses ranging from 4th- to 50th-orders transcends the previously reported second-/third-order nonlinear electric transports, and evidences *e*-HHG unequivocally. The observation of *e*-HHG larger than 10th-order in bilayer graphene with an extremely small driving electric field (e.g., ~10⁻⁷ V/nm) is quite surprising, considering that the *o*-HHG typically requires a pump electric field more than seven orders of magnitude larger (e.g., > 1 V/nm)[2,3]. In addition, thanks to the broken inversion symmetry under an external displacement field, *e*-HHGs with both even- and odd-orders are present. This is different from *o*-HHG in atomic gases with only odd-orders[1-3], while analogy to *o*-HHG from a non-centrosymmetric monolayer MoS$_2$ crystal[9].

To further explore the characteristics of *e*-HHG, we probed the harmonic transport responses of bilayer graphene by standard lock-in techniques. Figure 2b plots the normalized even-order harmonic transport responses $V_{xy}^{m\omega}$ ranging from 2nd- to 50th-orders against *n*, measured at $I_0 = 50$ nA and $D = 0.3$ V/nm. Please also refer to Supplementary Note 4 for odd-order results ranging from 3rd- to 49th-orders. Significant harmonic voltage responses exist at all orders. This is consistent with the spectrum results shown in Fig. 2a and further confirms the *e*-HHG in bilayer graphene. Figure 2c presents the maximum field conversion efficiencies, defined as $\eta = \frac{V_{xy}^{m\omega}}{V_{xx}^{\omega}} \cdot \frac{L}{W}$, as a function of *e*-HHG orders[4,43]. The maximum field conversion efficiencies decay with increasing the *e*-HHG orders and can be well fitted by $\eta \propto m^{-p}$ with $p = 2.02 \pm 0.05$. Such a power law scaling suggests the non-perturbative nature and has also been observed previously in *o*-HHG[44,45]. Remarkably, the field conversion efficiencies of *e*-HHG in bilayer graphene are extremely large. For example, the field conversion efficiency of 50th-order *e*-HHG can reach $5.6 \times 10^{-3}$, a value comparable to that of the state-of-the-art optical third-harmonic generation in graphene, and beyond that of typical 50th-order *o*-HHG by more than three orders of magnitude[5,43,45,46]. Such extremely efficient *e*-HHG in bilayer graphene facilitates to observe much higher-order *e*-HHGs, which we shall come to shortly.

**Non-perturbative nature of *e*-HHG**

Generally, the *m*-order *e*-HHG response $V_{xy}^{m\omega}$ can be expressed by the driving electric field $E^{\omega}$ as $V_{xy}^{m\omega} = \sigma_{xy}^{m\omega}(E^{\omega})^{\alpha}$. In the perturbative framework, $V_{xy}^{m\omega}$ should be proportional to the *m*-th power of $E^{\omega}$, i.e., $\alpha = m$. By contrast, the *m*-order *e*-HHG response is in the non-perturbative regime when $\alpha < m$ [4]. To reveal whether the observed *e*-HHG in bilayer graphene is perturbative or non-perturbative, we measured the high-order harmonic transport responses $V_{xy}^{m\omega}$ against the longitudinal $V_{xx}^{\omega}$. Figure 2d shows the results for $m = 4, 5, 6, 7, 8$. Power-law fits to data (solid lines in Fig. 2d) yield $V_{xy}^{4\omega} \propto (V_{xx}^{\omega})^{3.19\pm0.02}$, $V_{xy}^{5\omega} = (V_{xx}^{\omega})^{3.53\pm0.03}$, $V_{xy}^{6\omega} = (V_{xx}^{\omega})^{3.64\pm0.02}$, $V_{xy}^{7\omega} = (V_{xx}^{\omega})^{4.19\pm0.04}$, and $V_{xy}^{8\omega} = (V_{xx}^{\omega})^{4.27\pm0.02}$. These results clearly show a saturation-like behaviour and largely deviate from the character in the perturbative limit, that is, $m$th-order nonlinear harmonic response $V_{xy}^{m\omega}$ should scale as $(V_{xx}^{\omega})^m$ (dashed red lines in Fig. 2d). Consequently, e-HHG in bilayer graphene can advance to the non-perturbative and extreme nonlinear regime, in close resemblance to the o-HHG previously reported[3,4,9,47]. Further, e-HHG response also has a phase shift with respect to the driving current (Supplementary Note 5), similar to the chirp-delay effect observed in o-HHG[5].

**e-HHG up to 300th-order**

Figure 3 presents the 100th- (upper panel), 200th- (middle panel), and 300th-order (lower panel) e-HHG responses for a graphite-contacted bilayer graphene device at $D = 0.3$ V/nm (Supplementary Fig. 7). The ultrahigh-order e-HHG signals have been repeatedly measured three times to confirm their validity. These results undoubtedly shed light on the fascinating e-HHG up to 300th-order, which is also confirmed in a Cr/Au-contacted bilayer graphene device (Supplementary Fig. 8). In fact, non-vanishing 400th-order e-HHG is also reachable (Supplementary Fig. 9), but it has almost approached the limit of our measurement setup. We highlight that the 300th-order e-HHG observed here is the record-high order nonlinear responses achieved thus far in the condensed matter, much larger than the state-of-the-art order of o-HHG in solids (< 50th-order)[2,3]. Further, a much higher-order e-HHG comparable to the highest order of o-HHG in atomic gases (greater than 5000) may be expected by engineering nonlinear susceptibilities with external degrees of freedom such as magnetic field, high-pressure, and moiré engineering[48,49].

**Oscillations of e-HHG**

Fairly surprisingly, e-HHGs above 10th-order exhibit clear oscillations with the driving currents and the oscillations become even more prominent as the harmonic order increases, in striking contrast to the monotonically increased voltage responses of second- and third-order harmonic transports (Fig. 4a). The oscillation nature of e-HHGs deviates radically from the behaviour for a perturbative nonlinear response, and further establishes the extreme non-perturbative nature of the generation process. It is noteworthy that although non-perturbative features of high-order nonlinear effects have been unraveled in previous o-HHG studies, they are limited to only saturation-like behaviour[2,3]. This is the first time that the oscillations of high-order nonlinear effects with driving electric fields are observed.

Figure 4b presents the 200th-order e-HHG response $V_{xy}^{200\omega}$ as a function of the reciprocal of the driving current, i.e., $1/I_0$, under $n = 3 \times 10^{10}\ cm^{-2}$ and $D = 0.3$ V/nm. Obviously, $V_{xy}^{200\omega}$ oscillates periodically with $1/I_0$ and the oscillation amplitude increases with the drive currents. Phenomenologically, $V_{xy}^{200\omega}$ can be well described by the following expression (red line):
$$V_{xy}^{200\omega} \propto e^{-\alpha/I_0} \sin\left[2\pi(I_f/I_0 + \theta)\right] \qquad (1)$$
The right first term in equation (1) indicates the exponential modulation of the oscillation amplitude with driving currents. The right second term in equation (1) represents the periodic oscillation in $1/I_0$, where $I_f$ is the oscillation frequency. For 200th-order e-HHG, the extracted $I_f$ is about $211.86 \pm 1.15\ nA$, which is further confirmed by fast Fourier transform (FFT) spectrum (inset of Fig. 4b). Figure 4c shows the oscillation frequency $I_f$ fitted by equation (1) as a function of the e-

HHGs orders. Interestingly, $I_f$ scales linearly with the *e*-HHGs orders. The periodic oscillations of the *e*-HHGs and the linear scaling of $I_f$ largely transcend the current understanding of high-order nonlinear effects.

Although extrinsic factors such as Schottky contact diode may induce *e*-HHG signal, we confirm that the observed *e*-HHG responses are intrinsic, rather than from the extrinsic possibilities. First, for graphite-contacted bilayer graphene devices with higher quality, *e*-HHG responses with stronger signals and better signal-to-noise ratios are observed, excluding the extrinsic possibilities such as contact effects. Second, the observed *e*-HHGs, including the anomalous oscillation with respect to $1/I_0$, the linear scaling of *e*-HHG order with $I_f$, and the power law scaling of $\eta$, are corroborated well by theoretical calculations which we shall come to shortly, demonstrating the intrinsic contributions. Please refer to Supplementary Note 6 for detailed discussions on the intrinsic nature of the observed *e*-HHG. Furthermore, *e*-HHGs have been confirmed under high-frequency excitation with a driving frequency up to ~10 MHz (Supplementary Note 7).

**Theoretical simulations of *e*-HHG**

To unveil the possible underlying physical mechanism of unconventional *e*-HHGs responses in the non-perturbative regime, we propose a heuristic model based on the singular distribution of Berry curvature near band edges. Please refer to Supplementary Note 8 for the model details. Specifically, Fig. 4d depicts the numerical results of the $m$-th order *e*-HHG response $V_{xy}^{m\omega}$ as a function of the driving current $I_0$, with the amplitude of $m$-th order signal amplified by a factor of $m^2$. Intriguingly, the calculated *e*-HHGs start to oscillate when the driving current exceeds a critical value, bearing striking resemblance to the experimental observations shown in Fig. 4a. Such an oscillating behaviour indicates the non-perturbative nature of the process. Figure 4e presents the dependence of the calculated 200th-order harmonic response against $1/I_0$. The spacing between successive peaks is identical, demonstrating the periodic oscillations with $1/I_0$. The calculated FFT spectrum in inset of Fig. 4e shows the oscillation frequency $I_f = 200\,nA$, in good quantitative agreement with experimental result of $I_f = 211.86 \pm 1.15\,nA$ (Fig. 4b). Furthermore, the calculated $I_f$ shows a linear dependence with respect to the *e*-HHG orders (Fig. 4f), corroborating well with the result in Fig. 4c. In addition, calculated $V_{xy}^{m\omega}$ as a function of the *e*-HHG order (Supplementary Fig. 16) also uncover a power law scaling of $V_{xy}^{m\omega} = m^{-p}$, where $p = 2.06$, matching with experimental result of $p = 2.02 \pm 0.05$ (Fig. 2c). The quantitative consistency between the theoretically simulated results and the experimental data suggests that the non-perturbative *e*-HHG responses may origin from the singular distribution of Berry curvature and drastic change of electronic state distribution near band edges. Further, we became aware of an independent theoretical work that also confirms the universality of the *e*-HHGs in bilayer graphene[50].

In conclusion, we demonstrate, for the first time, the *e*-HHG in high-quality, *h*-BN encapsulated natural bilayer graphene devices and twisted graphene moiré superlattices. In marked contrast to the second-/third-order harmonic transports that show the perturbative behaviour, *e*-HHG can advance to the exotic non-perturbative and extreme nonlinear regime. Furthermore, *e*-HHG above 10th-order exhibits periodic oscillation with the reciprocal of driving current, and the oscillation frequency scales linearly with the *e*-HHG orders. Theoretical simulations corroborate the experiments, highlighting the underlying physical mechanism of unconventional *e*-HHG responses in the non-perturbative regime. The non-perturbative, electrically tunable *e*-HHG up to 300th-order demonstrated here not only enriches significantly our knowledge of high-order nonlinear physics, but also opens unprecedented possibilities of exploring the extreme non-equilibrium quantum phenomena, intriguing band topologic/geometric properties, novel nanoelectronics and energy devices that are not accessible at lower orders.

**Methods**

**Fabrication of graphene devices.** High quality, *h*-BN encapsulated, dual-gated natural bilayer graphene devices are fabricated via a dry transfer and layer-by-layer assembly technique. First, bilayer graphene and suitable *h*-BN flakes (about 15-30 nm thick) were mechanically exfoliated on separate Si/SiO$_2$ (300 nm thick) substrates. The thickness of bilayer graphene and *h*-BN was determined through optical microscopy, Raman spectra, and atomic force microscopy. Then, we used poly (Bisphenol A carbonate) (PC) supported by polydimethylsiloxane (PDMS) on a glass slide to pick up *h*-BN, bilayer graphene, another *h*-BN flakes, and graphite on the Si/SiO$_2$ substrates in sequence. Subsequently, the *h*-BN encapsulated bilayer graphene stacks were deposited onto a substrate. The fabrication of the metal top-gate and electrodes followed a standard electron-beam lithography process and electron-beam metal evaporation. The devices were designed in Hall bar structure and shaped by reactive ion etching (RIE) with a CHF$_3$ and O$_2$ gas mixture. Finally, all the bars were contacted via a 1D edge contact with Cr/Au electrodes (3 nm/30 nm). The twisted graphene devices were also fabricated in a similar way. A home-made micro-position stage and a typical 'tear and stack' technique were used to control the rotation angle.

**Nonlinear harmonic transport measurements.** Electrical transport was measured in a cryostat with a base temperature of 1.7 K. The top and bottom gate voltages were applied through two Keithley 2400 SourceMeters. Nonlinear harmonic transport was measured by standard lock-in techniques. Briefly, a sinusoidal alternating current $I^\omega = I_0 \sin(\omega t)$ with amplitude $I_0$ and oscillating frequency *ω* of 10–200 Hz was applied to the devices longitudinally, a transverse voltage $V_{xy}^{m\omega}$ with frequency *mω* was collected by a lock-in amplifier (Stanford Research Systems Model SR830), where the integer *m* (≥2) denotes the order of nonlinear harmonic transport. A big resistor of 10Mohm is connected in series with the sample to maintain a constant current output. The frequency and time-domain spectra of nonlinear electric transport were acquired directly by a digital oscilloscope.

**Nonlinear conductivity.** In general, the current density of *m*th-order *e*-HHG driven by an electric field $E_{xx}^\omega$ with oscillating frequency *ω* can be written as $j_{xy}^{m\omega} = \sigma_{xy}^{m\omega}(E_{xx}^\omega)^m$, where $\sigma_{xy}^{m\omega}$ is the *m*th-order nonlinear conductivity. In addition, the *mω* current response generates the electric field of frequency *mω*, given by $j_{xy}^{m\omega} = \sigma_0 E_{xy}^{m\omega}$, where $\sigma_0$ is linear conductivity. Considering that longitudinal voltage $V_{xx}^\omega = E_{xx}^\omega L$ and *m*th-order *e*-HHG voltage $V_{xy}^{m\omega} = E_{xy}^{m\omega} W$, we obtain *m*th-order nonlinear conductivity $\sigma_{xy}^{m\omega} = \sigma_0 \frac{V_{xy}^{m\omega} L^m}{(V_{xx}^\omega)^m W}$, where *L* (*W*) is the device channel length (channel width).

**Acknowledgements**
We thank Xiaobo Lu and Zuo Feng for providing graphite-contacted bilayer graphene samples and are grateful for the valuable discussions with Xi Dai and Yuelin Shao. We acknowledge supports from National Natural Science Foundation of China (NSFC, Grant Nos. 61888102, 12274447), the National Key Research and Development Program of China (Grant Nos. 2021YFA1202900, 2020YFA0309600, 2023YFA1407000, 2021YFA1400502), the Strategic Priority Research Program of CAS (Grant Nos. XDB0470101) and the Key-Area Research and Development Program of Guangdong Province (Grant No. 2020B0101340001). K.W. and T.T. acknowledge support from the JSPS KAKENHI (Grant Numbers 20H00354, 21H05233 and 23H02052) and World Premier International Research Center Initiative (WPI), MEXT, Japan.

**Author contributions**

G.Z. and L.D. proposed and supervised the research project. L.D., X.Z., W.Y., and G.Z. designed the experiments and analyzed the data; X.Z. fabricated the bilayer and twisted trilayer graphene devices, performed the *e*-HHG measurements under the supervision of L.D., W.Y. and G.Z.; Z.P. and W.Y. built the high-frequency set-up and performed the *e*-HHG measurements in the MHz regime; J. D., and J.Z. assisted in the preparation of monolayer graphene devices and measurement; L.L. and Y.C. fabricated the twisted graphene devices; K.W. and T.T. provided *h*-BN crystals; X.X. proposed the theory, M.G. and H.L. performed the theoretical simulations; L.D., G.Z., X.Z., M.G., and H.L. wrote the manuscript with the input from all authors.

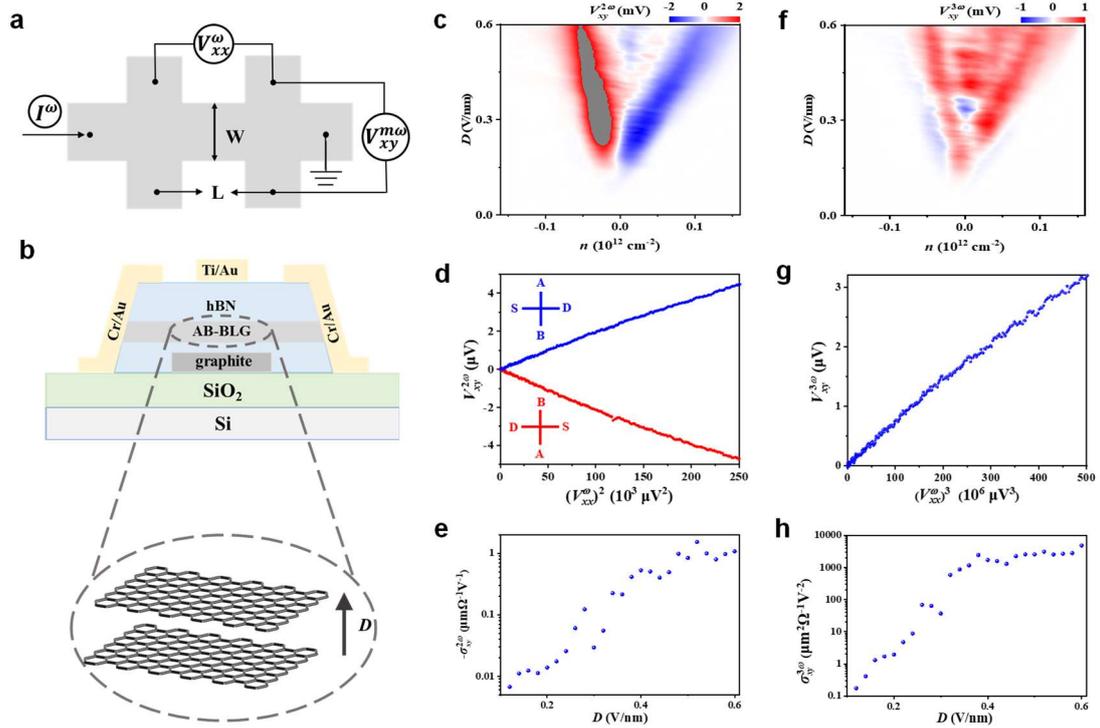

**Fig. 1 | Electrically tunable quantum nonlinearity in bilayer graphene. a,** Schematic of nonlinear harmonic transport measurements along transverse direction in a standard Hall bar device. $L/W$ denotes the device channel length/width. **b**, Top: illustration of a $h$-BN encapsulated, dual-gated bilayer graphene device, with graphite (Au) as the bottom (top) gate. Bottom: close-up of bilayer graphene structure. **c,f,** Colour plot of second-order (**c**) and third-order (**f**) harmonic voltage responses against the displacement field $D$ and carrier density $n$. Inset in (**c**) is the colour map of linear longitudinal resistance $R_{xx}$ as a function of $D$ and $n$. **d,g,** $V_{xy}^{2\omega}$ (**d**) and $V_{xy}^{3\omega}$ (**g**) as a function of the square and cube of the linear longitudinal voltage response $V_{xx}^{\omega}$, respectively. The insets in (**d**) show the electrode geometry for measurements. The current is injected from the 'S' to the 'D' electrode and the voltage is measured between the 'A' and 'B' electrodes. **e,h,** Second-order (**e**) and third-order (**h**) nonlinear conductivity versus the displacement field $D$.

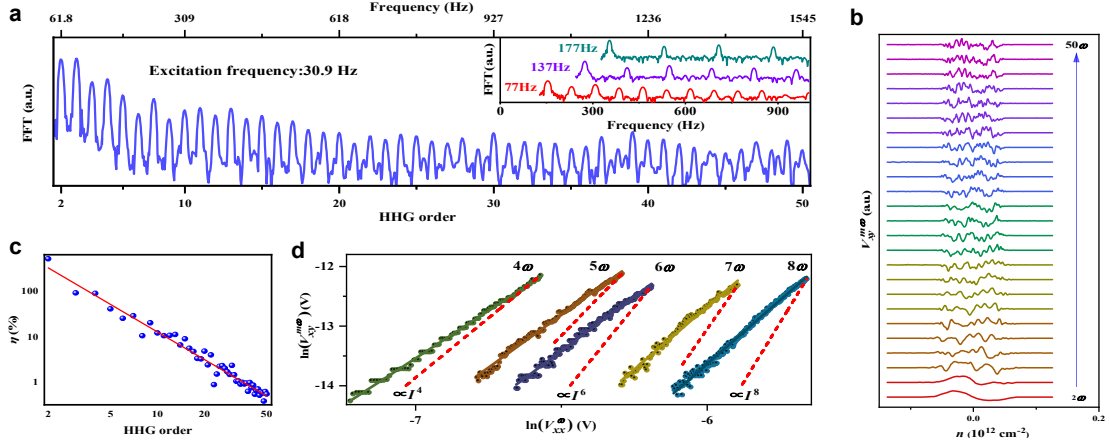

**Fig. 2 | *e*-HHG in non-perturbative regime. a,** Frequency spectrum excited by an alternating current with amplitude of 500 nA and oscillating frequency of 30.9 Hz. Inset: frequency spectra under different driving frequencies: 77 Hz (red), 137 Hz (gray) and 177 Hz (cyan). $D = 0.3$ V/nm and $n = 3.5 \times 10^{10}$ $cm^{-2}$. **b,** Normalized even-order response $V_{xy}^{m\omega}$ ranging from 2nd- to 50th-order against $n$ measured at $I_0 = 50$ nA and $D = 0.3$ V/nm. Offset is set for better resolution. **c,** The maximum field conversion efficiencies, defined as $\eta = \frac{V_{xy}^{m\omega}}{V_{xx}^{\omega}} \cdot \frac{L}{W}$, as a function of the *e*-HHG orders. **d,** The measured nonlinear harmonic response $V_{xy}^{m\omega}$ as a function of the longitudinal $V_{xx}^{\omega}$ for harmonic orders ranging from 4th- to 8th-order (dots). A fit of the experimental data to a power law, $(V_{xx}^{\omega})^q$, yields a phenomenological exponent of $q = 3.19 \pm 0.02$ for 4th-order, $q = 3.53 \pm 0.03$ for 5th-order, $q = 3.64 \pm 0.02$ for 6th-order, $q = 4.19 \pm 0.04$ for 7th-order, and $q = 4.27 \pm 0.02$ for 8th-order (solid lines). This strongly deviates from the behaviour of a perturbative nonlinear response of $q = m$ (red dashed lines).

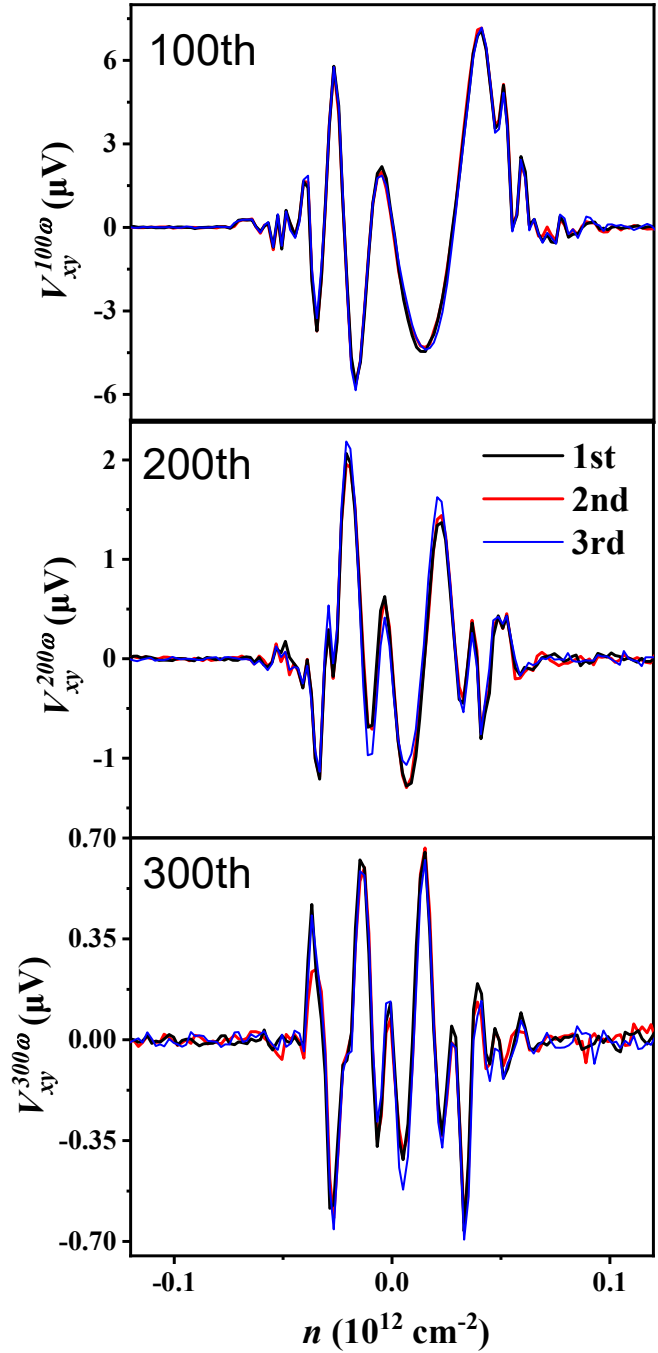

**Fig. 3 | *e*-HHG up to 300th-order.** *e*-HHG response $V_{xy}^{m\omega}$ as a function of $n$ at $D = 0.3$ V/nm for representative harmonic orders $m = 100$ (upper panel), 200 (middle panel), and 300 (lower panel). For the measured ultrahigh-order signals, we have repeated three times to confirm their validity.

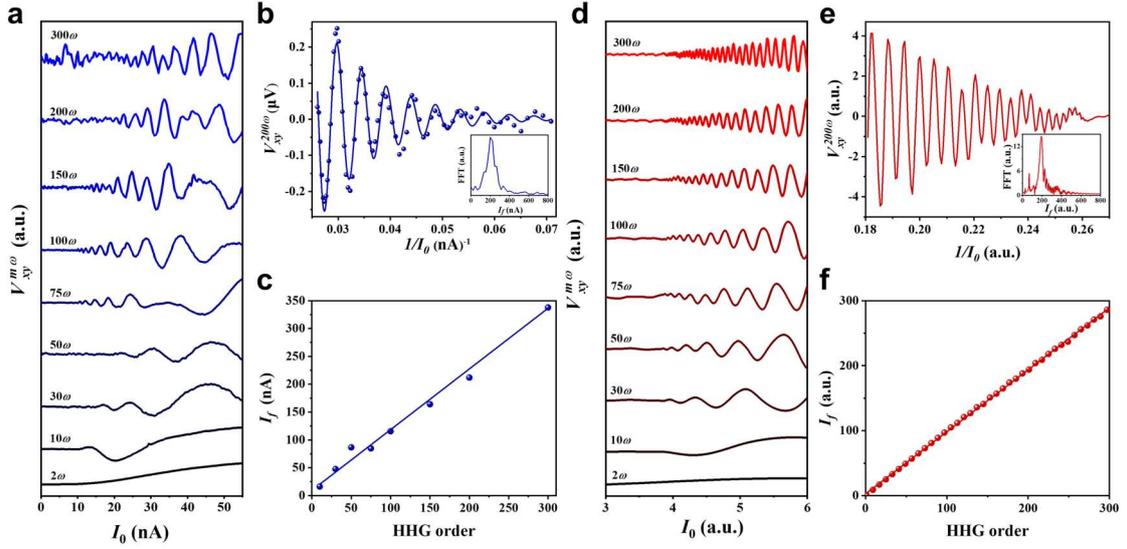

**Fig. 4 | Oscillations of *e*-HHG. a,** Nonlinear harmonic transport response $V_{xy}^{m\omega}$ as a function of the driving currents. The *e*-HHGs above 10th-order exhibit oscillations with the driving currents and the oscillations become more pronounced as the order increases. **b,** The 200th-order *e*-HHG response $V_{xy}^{200\omega}$ against the reciprocal of the driving current, i.e., $1/I_0$. Red line is the fitting result with $V_{xy}^{200\omega} \propto e^{-\alpha/I_0}\sin\left[2\pi(I_f/I_0 + \theta)\right]$. Inset: fast Fourier transform (FFT) spectrum. **c,** Oscillation frequency $I_f$ depends linearly with the *e*-HHG order. $n = 3 \times 10^{10}\ cm^{-2}$ and $D = 0.3$ V/nm. **d-f** are the corresponding theoretical simulations.

1   Winterfeldt, C., Spielmann, C. & Gerber, G. Colloquium: Optimal control of high-harmonic generation. *Rev. Mod. Phys.* **80**, 117-140, doi:10.1103/RevModPhys.80.117 (2008).

2   Ghimire, S. & Reis, D. A. High-harmonic generation from solids. *Nat. Phys.* **15**, 10-16, doi:10.1038/s41567-018-0315-5 (2019).

3   Goulielmakis, E. & Brabec, T. High harmonic generation in condensed matter. *Nat. Photon.* **16**, 411-421, doi:10.1038/s41566-022-00988-y (2022).

4   Boyd, R. W. *Nonlinear optics*.   (Academic press, 2020).

5   Krausz, F. & Ivanov, M. Attosecond physics. *Rev. Mod. Phys.* **81**, 163-234, doi:10.1103/RevModPhys.81.163 (2009).

6   Ghimire, S. *et al.* Observation of high-order harmonic generation in a bulk crystal. *Nat. Phys.* **7**, 138-141, doi:10.1038/nphys1847 (2011).

7   Popmintchev, D. *et al.* Ultraviolet surprise: Efficient soft x-ray high-harmonic generation in multiply ionized plasmas. *Science* **350**, 1225-1231, doi:doi:10.1126/science.aac9755 (2015).

8   Luu, T. T. *et al.* Extreme ultraviolet high-harmonic spectroscopy of solids. *Nature* **521**, 498-502, doi:10.1038/nature14456 (2015).

9   Liu, H. *et al.* High-harmonic generation from an atomically thin semiconductor. *Nat. Phys.* **13**, 262-265, doi:10.1038/nphys3946 (2017).

10  Vampa, G., Fattahi, H., Vučković, J. & Krausz, F. Attosecond nanophotonics. *Nat. Photon.* **11**, 210-212, doi:10.1038/nphoton.2017.41 (2017).

11  Garg, M. *et al.* Multi-petahertz electronic metrology. *Nature* **538**, 359-363, doi:10.1038/nature19821 (2016).

12  Goulielmakis, E. *et al.* Attosecond Control and Measurement: Lightwave Electronics. *Science* **317**, 769-775, doi:10.1126/science.1142855 (2007).

13  Sodemann, I. & Fu, L. Quantum Nonlinear Hall Effect Induced by Berry Curvature Dipole in Time-Reversal Invariant Materials. *Phys. Rev. Lett.* **115**, 216806, doi:10.1103/PhysRevLett.115.216806 (2015).

14  Ma, Q. *et al.* Observation of the nonlinear Hall effect under time-reversal-symmetric conditions. *Nature* **565**, 337-342, doi:10.1038/s41586-018-0807-6 (2019).

15  Kang, K., Li, T., Sohn, E., Shan, J. & Mak, K. F. Nonlinear anomalous Hall effect in few-layer WTe$_2$. *Nat. Mater.* **18**, 324-328, doi:10.1038/s41563-019-0294-7 (2019).

16  Du, Z. Z., Lu, H.-Z. & Xie, X. C. Nonlinear Hall effects. *Nat. Rev. Phys.* **3**, 744-752, doi:10.1038/s42254-021-00359-6 (2021).

17  Ideue, T. & Iwasa, Y. Symmetry Breaking and Nonlinear Electric Transport in van der Waals Nanostructures. *Annu. Rev. Condens. Matter Phys.* **12**, 201-223, doi:10.1146/annurev-conmatphys-060220-100347 (2021).

18  Juergens, P. *et al.* Linking High-Harmonic Generation and Strong-Field Ionization in Bulk Crystals. *ACS Photon.* **11**, 247-256, doi:10.1021/acsphotonics.3c01436 (2024).

19  Lai, S. *et al.* Third-order nonlinear Hall effect induced by the Berry-connection polarizability tensor. *Nat. Nanotechnol.* **16**, 869-873, doi:10.1038/s41565-021-00917-0 (2021).

20  He, P. *et al.* Graphene moiré superlattices with giant quantum nonlinearity of chiral Bloch electrons. *Nat. Nanotechnol.* **17**, 378-383, doi:10.1038/s41565-021-01060-6 (2022).

21  Kumar, D. *et al.* Room-temperature nonlinear Hall effect and wireless radiofrequency

| | |
|---|---|
| | rectification in Weyl semimetal TaIrTe₄. *Nat. Nanotechnol.* **16**, 421-425, doi:10.1038/s41565-020-00839-3 (2021). |
| 22 | Ho, S.-C. *et al.* Hall effects in artificially corrugated bilayer graphene without breaking time-reversal symmetry. *Nat. Electron.* **4**, 116-125, doi:10.1038/s41928-021-00537-5 (2021). |
| 23 | Ando, F. *et al.* Observation of superconducting diode effect. *Nature* **584**, 373-376, doi:10.1038/s41586-020-2590-4 (2020). |
| 24 | Wakatsuki, R. *et al.* Nonreciprocal charge transport in noncentrosymmetric superconductors. *Sci. Adv.* **3**, e1602390, doi:doi:10.1126/sciadv.1602390 (2017). |
| 25 | Avci, C. O. *et al.* Current-induced switching in a magnetic insulator. *Nat. Mater.* **16**, 309-314, doi:10.1038/nmat4812 (2017). |
| 26 | Ideue, T. *et al.* Bulk rectification effect in a polar semiconductor. *Nat. Phys.* **13**, 578-583, doi:10.1038/nphys4056 (2017). |
| 27 | Itahashi, Y. M. *et al.* Giant second harmonic transport under time-reversal symmetry in a trigonal superconductor. *Nat. Commun.* **13**, 1659, doi:10.1038/s41467-022-29314-4 (2022). |
| 28 | Kang, K. *et al.* Switchable moiré potentials in ferroelectric $WTe_2/WSe_2$ superlattices. *Nat. Nanotechnol.*, doi:10.1038/s41565-023-01376-5 (2023). |
| 29 | Lin, J.-X. *et al.* Spontaneous momentum polarization and diodicity in Bernal bilayer graphene. *arXiv:2302.04261* (2023). |
| 30 | Gao, A. *et al.* Quantum metric nonlinear Hall effect in a topological antiferromagnetic heterostructure. *Science* **381**, 181-186, doi:10.1126/science.adf1506 (2023). |
| 31 | Wang, N. *et al.* Quantum-metric-induced nonlinear transport in a topological antiferromagnet. *Nature* **621**, 487-492, doi:10.1038/s41586-023-06363-3 (2023). |
| 32 | Duan, J. *et al.* Giant Second-Order Nonlinear Hall Effect in Twisted Bilayer Graphene. *Phys. Rev. Lett.* **129**, 186801, doi:10.1103/PhysRevLett.129.186801 (2022). |
| 33 | Huang, M. *et al.* Giant nonlinear Hall effect in twisted $WSe_2$. *Natl. Sci. Rev.* **10**, nwac232, doi:10.1093/nsr/nwac232 (2022). |
| 34 | Sinha, S. *et al.* Berry curvature dipole senses topological transition in a moiré superlattice. *Nat. Phys.* **18**, 765-770, doi:10.1038/s41567-022-01606-y (2022). |
| 35 | Du, L. *et al.* Engineering symmetry breaking in 2D layered materials. *Nat. Rev. Phys.* **3**, 193-206 (2021). |
| 36 | Huang, M. *et al.* Intrinsic Nonlinear Hall Effect and Gate-Switchable Berry Curvature Sliding in Twisted Bilayer Graphene. *Phys. Rev. Lett.* **131**, 066301 (2023). |
| 37 | Liu, H. *et al.* Berry connection polarizability tensor and third-order Hall effect. *Phys. Rev. B* **105**, 045118, doi:10.1103/PhysRevB.105.045118 (2022). |
| 38 | Cheng, Y. *et al.* Third harmonic characterization of antiferromagnetic heterostructures. *Nat. Commun.* **13**, 3659, doi:10.1038/s41467-022-31451-9 (2022). |
| 39 | Zhang, Y. *et al.* Direct observation of a widely tunable bandgap in bilayer graphene. *Nature* **459**, 820-823, doi:10.1038/nature08105 (2009). |
| 40 | Isobe, H., Xu, S.-Y. & Fu, L. High-frequency rectification via chiral Bloch electrons. *Sci. Adv.* **6**, eaay2497, doi:doi:10.1126/sciadv.aay2497 (2020). |
| 41 | Sui, M. *et al.* Gate-tunable topological valley transport in bilayer graphene. *Nat. Phys.* **11**, 1027-1031, doi:10.1038/nphys3485 (2015). |

42   Shimazaki, Y. *et al.* Generation and detection of pure valley current by electrically induced Berry curvature in bilayer graphene. *Nat. Phys.* **11**, 1032-1036, doi:10.1038/nphys3551 (2015).

43   Teubner, U. & Gibbon, P. High-order harmonics from laser-irradiated plasma surfaces. *Rev. Mod. Phys.* **81**, 445-479, doi:10.1103/RevModPhys.81.445 (2009).

44   Norreys, P. A. *et al.* Efficient Extreme UV Harmonics Generated from Picosecond Laser Pulse Interactions with Solid Targets. *Phys. Rev. Lett.* **76**, 1832-1835, doi:10.1103/PhysRevLett.76.1832 (1996).

45   Dromey, B. *et al.* High harmonic generation in the relativistic limit. *Nat. Phys.* **2**, 456-459, doi:10.1038/nphys338 (2006).

46   Hafez, H. A. *et al.* Extremely efficient terahertz high-harmonic generation in graphene by hot Dirac fermions. *Nature* **561**, 507-511, doi:10.1038/s41586-018-0508-1 (2018).

47   Yoshikawa, N., Tamaya, T. & Tanaka, K. High-harmonic generation in graphene enhanced by elliptically polarized light excitation. *Science* **356**, 736-738, doi:doi:10.1126/science.aam8861 (2017).

48   Popmintchev, T. *et al.* Bright Coherent Ultrahigh Harmonics in the keV X-ray Regime from Mid-Infrared Femtosecond Lasers. *Science* **336**, 1287-1291, doi:10.1126/science.1218497 (2012).

49   Du, L. *et al.* Moiré photonics and optoelectronics. *Science* **379**, eadg0014, doi:10.1126/science.adg0014 (2023).

50   Shao, Y. & Dai, X. J. a. p. a. Giant High-order Nonlinear and Nonreciprocal Electrical Transports Induced by Valley Flipping in Bernal Bilayer Graphene. *arXiv:2403.19498* (2024).